\documentclass[aps,twocolumn,nofootinbib,showpacs,floatfix]{revtex4}
\usepackage{graphics} 
\usepackage{amssymb,amsmath}
\usepackage{amsmath}
\usepackage{psfrag}
\usepackage{graphicx}
\usepackage{wrapfig}

\def\l|an{\left\langle}

\def\be{\begin{equation}} 
\def\ee{\end{equation}} 
\def\bea{\begin{eqnarray}} 
\def\eea{\end{eqnarray}}  
\def\bean{\begin{eqnarray*}} 
\def\eean{\end{eqnarray*}}

\def\bk{{\bf k}}  
\def\bx{{\bf x}}

\def\ba{{\bf a}} 
 
\def\bFo{{\bf F}} 
\def\bfo{{\bf f}}

\def\bu{{\bf u}}

\def\by{{\bf y}}

\def\bq{{\mathbf q}}
\def\bse{\begin{subequations}}
\def\ese{\end{subequations}}

\def\bF{{\mathbf F}}

\def\bq{\mathbf{q}}

\def\lsim{\raise 0.4ex\hbox{$<$}\kern -0.8em\lower 0.62ex\hbox{$\sim$}} 
\def\gsim{\raise 0.4ex\hbox{$>$}\kern -0.7em\lower 0.62ex\hbox{$\sim$}}

\def\bk{\mathbf{k}}

\def\f0N{f_0^{(N)}}
\def\bec{\begin{center}}
\def\eec{\end{center}}

\begin{document} 
\title{A dynamical classification of the range of pair interactions}
\author{A. Gabrielli$^{1,2}$, M. Joyce$^{3,4}$, B. Marcos$^{5}$ and F. Sicard$^{3}$} 
\affiliation{$^1$SMC, CNR-INFM, Physics Department, University 
``Sapienza'' of Rome, Piazzale Aldo Moro 2, 00185-Rome, Italy}
\affiliation{$^2$Istituto dei Sistemi Complessi - CNR, Via dei Taurini 19, 
00185-Rome, Italy}
\affiliation{$^3$Laboratoire de Physique Nucl\'eaire et Hautes \'Energies,\\
Universit\'e Pierre et Marie Curie - Paris 6,
CNRS IN2P3 UMR 7585, 4 Place Jussieu, 75752 Paris Cedex 05, France}
\affiliation{$^4$Laboratoire de Physique Th\'eorique de la Mati\`ere Condens\'ee,\\
Universit\'e Pierre et Marie Curie - Paris 6,
CNRS UMR 7600, 4 Place Jussieu, 75752 Paris Cedex 05, France}
\affiliation{$^5$ Laboratoire J.-A. Dieudonn\'e, UMR 6621, 
Universit\'e de Nice --- Sophia Antipolis,
Parc Valrose 06108 Nice Cedex 02, France} 

\begin{abstract}   
\begin{center}    
{\large\bf Abstract}
\end{center}    
We formalize 
a classification of pair interactions based on the convergence properties
of the {\it forces} acting on particles as a function of system size. We do so 
by considering the behavior of the  probability distribution function (PDF) $P(\bF)$ of the force field $\bF$ 
in a particle distribution in the limit that the size of the system is taken
to infinity at constant particle density, i.e., in the ``usual'' thermodynamic  limit. 
For a pair interaction potential $V(r)$ with $V(r \rightarrow \infty) \sim 1/r^\gamma$
defining a {\it bounded} pair force, we show that $P(\bF)$ converges 
continuously to a well-defined and rapidly decreasing PDF 
if and only if the {\it pair force} is absolutely integrable, i.e., 
for $\gamma > d-1$, where $d$ is the
spatial dimension. We refer to this case as {\it dynamically short-range}, 
because the dominant contribution to the force on a typical particle 
in this limit arises from particles in a finite neighborhood 
around it. For the {\it dynamically long-range} case, i.e., $\gamma \leq d-1$, 
on the other hand, the dominant contribution to the force comes from the 
mean field due to the bulk, which becomes  undefined in this  limit. 
We discuss also how,  for  $\gamma \leq d-1$ (and notably, for the case 
of gravity, $\gamma=d-2$) $P(\bF)$ may, in some cases, be defined in a 
weaker sense. This involves a regularization of the force summation
which is generalization of the procedure
employed to define gravitational forces in an infinite static homogeneous 
universe. We explain that the relevant classification in this context is, however,
that which divides pair forces with $\gamma >  d-2$ (or $\gamma < d-2$), 
for which the PDF of the  {\it difference in forces} is defined
(or not defined) in the infinite system limit, without any regularization.
In the former case dynamics can, as for the (marginal) case of gravity,  be 
defined  consistently in an infinite uniform system.
 \end{abstract}    
\maketitle   
\date{today}  

\twocolumngrid   

\section{introduction}

Interactions are traditionally classified as long-range 
(or short-range) with respect to the non-additivity 
(or additivity) of the {\it potential energy} in the usual 
thermodynamic limit, i.e., when the number of particles
$N$ and volume $V$ are taken to infinity at constant
particle density. This is the property which 
determines the way in which standard instruments
of statistical mechanics are applied to determine 
{\it equilibrium} properties (see e.g. \cite{ruelle, Dauxoisetal, assisi}). 
Indeed in the case of long-range interactions, these instruments 
are applied using an appropriately generalized 
thermodynamic limit, in which the coupling or density are also 
scaled with system size. Such an analysis gives rise
generically to features at equilibrium which are  
qualitatively different from those in short-range systems
 --- inhomogeneous statistical equilibria, non-equivalence 
of statistical ensembles, negative specific heat in the 
microcanonical ensemble (see e.g. \cite{Dauxoisetal, assisi}). 
Most of these unusual features were first noted and
studied in the context of the study of gravitating 
systems in astrophysics 
(see e.g. \cite{thirring_book, chavanis_phase+transitions_2006} 
for reviews), and it has been realized in recent years
that they are more generic in long-range interacting systems. 
This thermodynamic analysis extended to long-range systems 
is believed to determine, however, the behavior of such systems 
only on time scales which diverge as some power of $N$ (when 
expressed in terms of the characteristic
dynamical time scales). On shorter times scales --- usually those
of interest in practical applications --- study of several
such systems (see e.g. \cite{yawn+miller_2003, yamaguchi_etal_04, antoniazzi1, antoniazzi2, yamaguchi_1dgravLB_2008}
and references therein) shows that they appear generically, like in the 
well-documented case of gravity, to relax from almost
any initial conditions to (almost) time-independent states 
--- referred to variously as ``meta-equilibria'', ``quasi-equilibria'' 
or ``quasi-stationary states'' (QSS). The physics of these
states, which are generically very different from those at
thermal equilibrium,  is understood to be the result of evolution 
in the collisionless regime described by Vlasov equation 
(usually referred to as the ``collisionless Boltzmann equation'' 
in the astrophysical literature \cite{binney}). Both the genesis of these 
states and their long-time relaxation are poorly 
understood, and are the subject of active study (see e.g. 
\cite{campa_etal2007, Baldovin_etal2009, chavanis_kEqns_2010, teles_etal2010, gupta+mukamel_2010}).

In this article we consider a simple classification of pair
interactions different to this usual thermodynamic one.
Instead of considering the convergence properties of
{\it potential energy} in the usual thermodynamic limit, we consider
those of the {\it force}, in the same limit. 
The resulting classification can, like the usual one,
be understood easily from simple considerations.
To see this let us consider, as illustrated 
schematically  in Fig.~1,
\begin{figure}[h!]
\begin{center}
\includegraphics[width=7cm]{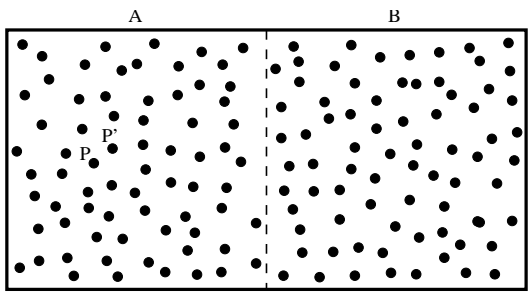}
\label{figure1}
\caption{An (approximately) uniform system consisting of two 
sub-systems A and B.} 
\end{center}
\end{figure}
 a uniform system of particles
interacting by a pair potential $V(r \to \infty) \sim 1/r^\gamma$,
and divided into two pieces, $A$ and $B$. For the usual
thermodynamic classification one can consider that when 
the potential $V(r)$ is integrable at large $r$, i.e., 
$\gamma > d$ where $d$ is the spatial dimension, the
potential energy of a typical particle comes essentially from
its interaction with particles in a finite region about it. 
The energy of a particle (e.g. $P$ or $P'$) in $A$ is thus insensitive to
whether $B$ is present or not (and thus the total energy 
is equal, up to surface effects, to the sum of the
energies of the subsystems): from an energetic point of 
view a particle  ``does not care''  what the size of the system
is, and the interaction is in this sense short-range.
The distinction we consider is the analogous one
deduced when one reasons in terms of {\it force}
(or acceleration) rather than potential energy, and since forces are the
primary physical quantities in dynamics, we refer
to the corresponding classification as one of  {\it dynamical} range. 
It is straightforward to see that in such a system,  
if the pair force is absolutely integrable, i.e., $\gamma > d-1$,
the force acting on a typical particle is due essentially
to its interaction with particles in a finite neighborhood
around it, while if  $\gamma < d-1$ this is not the case.
Thus in the former case a particle in $A$ ``does not care''
whether the sub-system $B$ is present or not, and 
in this sense the interaction is ``dynamically short-range".
The classification differs from the standard one for interactions 
with $d < \gamma < d-1$:  for such interactions the potential
energy ``sees the bulk'', but the force, which is its derivative,
does not.  

While the principle motivation for defining such a classification 
is that  it may be relevant to understanding the qualitative behaviors 
of the out of equilibrium dynamics of such systems, it is {\it not} the aim
of this article to establish that this is the case.  We will limit ourselves 
in this respect to some brief remarks in our conclusions below.
Our goal here is to provide a precise formulation of such a classification 
of the range of interactions based on the convergence properties of forces.
While in the usual thermodynamic classification case one considers (see e.g.\cite{ruelle}) 
the mathematical properties of essential functions describing systems at equilibrium in
the limit $N \to \infty$, $V \to \infty$ at fixed particle density
$n_0=N/V$ (i.e. the usual thermodynamic limit), we will consider the
behavior of functions characterising the forces in this same limit.
More specifically we consider, following an approach introduced by
Chandrasekhar for the case of gravity \cite{chandra43, book}, the
definedness of the probability distribution function (PDF) of the
force field in statistically homogeneous particle distributions as
the size of the system becomes arbitrarily large.  Such distributions
and this limit may be described mathematically using the language of 
stochastic point processes, considering the class of such processes 
which have a well defined positive mean density when the infinite
size limit is considered.  To avoid any confusion we will refer to the usual
thermodynamic limit in this context simply as the {\it infinite system
  limit}. Indeed the existence or non-existence of the quantities we
are studying in this limit has no direct relation here to the
determination of properties at thermal equilibrium.  Further, in the
context of the literature on long-range interactions the term
``thermodynamic limit" is now widely associated with the generalized
such limit, which involves adopting a different scaling of $V$ (or possibly
coupling constants) with $N$  (for a discussion see e.g. \cite{balescu}).

In this article we also discuss, in Sec.~\ref{Definedness of dynamics in the infinite system limit}, 
a further (and different) classification which can be given of pair interactions according to their
range. This is relevant when one addresses more generally, for any given pair interaction,
a question which arises for Newtonian gravity in a cosmological setting:  can a consistent 
dynamics be defined in an infinite system with non-zero density? 
A rigorous approach to the same question and the connection with the possibility of
defining a {\em statistical mechanical} state for the system has been developed in 
\cite{lanford1,lanford2} in the particular case of a short range non-negative pair potential with finite support.
Our conclusion, which generalizes a previous
discussion given by two of us in \cite{1dPoisson_2010}, is that the
answer to this question is that a necessary and sufficient condition for 
such a dynamics is not the integrability  of the pair force, but 
instead the integrability of its gradient. This means that one requires 
$\gamma \geq  d-2$, with gravity in any dimension (i.e. $\gamma=d-2$, the
interaction potential solving the appropriate $d$-dimensional Poisson 
equation) being the marginal case in which such an infinite system
limit may be defined. The reason is simply that, in an infinite system 
without any  preferred point (i.e. when this limit is defined respecting 
statistical translational invariance), the physically meaningful quantity is the relative 
position of particles (as there is no meaning to absolute position).
It is thus the convergence of relative forces on particles with system size which 
matters. In terms of the schema given above the distinction arises thus when 
one considers two close-by points (e.g. $P$ and $P'$ in Fig.~1) in sub-system $A$, say,
and asks whether their relative forces --- and thus relative motions --- depend
on the presence of $B$ or not (or, equivalently, on the size
of the system). The answer is that this difference of forces does
not essentially depend on $B$  if the gradient of the 
pair force is absolutely integrable, i.e., $\gamma > d-2$,  as
in this case this difference is dominated by the contribution from
particles in a finite neighborhood around them.  Thus for 
the case that $d-2 < \gamma \leq  d-1$ the forces acting
on two such particles become ill defined as the size of the system is
extended to infinity, but their difference remains finite. Indeed,
as has been discussed in  \cite{1dPoisson_2010} in the context
of gravity in one dimension,  the
diverging component of the force on a particle represents a 
force on their centre of mass, which has no physical relevance in 
an infinite system without a preferred origin.

The paper is organized as follows. In the next section we recall the
essential properties of stochastic point processes of relevance to our
considerations, and then consider the general analyticity properties
of the PDF of the total force at an arbitrary spatial point in such
a particle distribution. We show that, for any pair force
which is bounded, this PDF in the infinite volume limit 
is either well defined and rapidly decreasing, or else 
vanishes pointwise, i.e., the total force is an 
ill defined stochastic quantity. This means that it suffices, when
studying pair potentials with different possible behaviors
at large scales, 
to show that some chosen moment of the PDF 
converges to a finite value in this limit (or diverges) in order to
establish that the whole PDF itself is well-defined (or ill
defined). In Sec.~\ref{PDF-F-mean-variance} we give a general and
formal expression for the variance of the total force PDF in a generic
infinite uniform stochastic process in terms of the pair force
and the two-point correlation properties of the SPP.
From this we then deduce our principal result that the force PDF
exists strictly in the infinite system limit if and only if the pair
force is absolutely integrable at large separations (i.e. $\gamma \leq
d-1$) , while it can be defined only in a weaker sense, introducing a
regularization, when the pair force is not absolutely integrable. In
the following section we discuss the physical relevance of the use of
such a regularization, which is the generalization of a 
simple formulation given by Kiessling \cite{kiessling} of 
that originally introduced by Jeans for the case of 
gravity \cite{jeans}, often misleadingly referred to as 
the ``Jeans swindle" \cite{binney, kiessling}.
By analyzing the evolution of density perturbations in
an infinite system, we show that the physical relevance of such a 
regularization of the forces requires also a  constraint on the behavior 
of the PDF of total {\it force differences} as a function of system size. 
This leads to the conclusion that $\gamma \leq d-2$ is the necessary 
and sufficient condition in order for it to be possible to have a well 
defined infinite  system limit at constant density for dynamics under
a given pair interaction.  In the conclusions we review briefly the the
relation of our results to previous work in the literature, and 
comment a little more on the possible relevance of our principle classification of
interactions into {\it dynamically short-range} and {\it dynamically
  long-range} to the study of the out of equilibrium dynamics of
  such systems.

\section{The force PDF in uniform stochastic point processes:  general results}
\label{The force PDF in uniform stochastic point processes}

We first recall the definitions of some basic quantities used in the
statistical characterization of a stochastic point process  and 
define the total force PDF (see e.g.
\cite{book} for a detailed discussion). We then  derive some results 
on the analyticity properties of the latter quantity which we will exploit 
in deriving our central results in the next section.

\subsection{Stochastic point processes}

In order to study the properties of the force field in the infinite
system limit given by $N \rightarrow \infty$, $V\rightarrow \infty$
with fixed average density $n_0>0$ for a large scale uniform and
spatially homogeneous particle system, we generalize the approach
introduced by Chandrasekhar in \cite{chandra43} for the total
gravitational field in a homogeneous Poisson particle distribution to 
more general cases and spatial dimensions. To do so we need to 
characterize statistically point-particle distributions in this limit,
and we do this using the language of stochastic point processes (SPP).  
The microscopic number density of a
single realization of the process is \be
\label{density}
n({\bf x})=\sum_{i} \delta \left( {\bf x} - {\bf x}_i \right) \ee
where $\delta$ is the $d$-dimensional Dirac delta function, $\bx_i$ is
the position of the $i^{th}$ system particle and the sum runs over all
the particles of the system.  We will limit our discussion to particle
distributions in a euclidean $d-$dimensional space which are (i)
statistically translationally invariant (i.e. spatially homogeneous or
stationary) and (ii) large scale uniform in the infinite volume limit.
Property (i) means that the statistical properties around a given
spatial point of the particle distribution do not depend on the
location of the point. In other words the statistical weights of
two realizations of the point process, of which one is the rigidly
translated version of the other, are the same and do not depend on the
translation vector.  In particular this implies that the ensemble
average (i.e. average over the realizations of the SPP)
$\left<n(\bx)\right>$ of the microscopic number density takes a
constant value $n_0>0$ independent of $\bx$.  Moreover the two-point
correlation function of the microscopic density
$\left<n(\bx)n(\bx')\right>$ depends only on the vector distance
$\bx-\bx'$.  Feature (ii) means that the average particle number
fluctuation $\delta
N(R)=(\left<N^2(R)\right>-\left<N(R)\right>^2)^{1/2}$ in a sphere of
radius $R$ increases slower with $R$ than the average number
$\left<N(R)\right>_0V(R)$ with $R$, where $V(R)\propto R^d$ is the
volume of the $d-$dimensional sphere.  

Let us start by considering a
generic realization of the particle distribution in a finite volume
$V$ and let the total number of particles of the given realization be
$N$.  The particle positions $\bx_i$ are fully characterized 
statistically by the joint probability densty function (PDF) ${\cal P}_N (\{ \bx_i \})$
conditional to having $N$ particles in the realization ($\{ \bx_i \}$
indicates the set of positions of all system particles in the given
realization). As a simple, but paradigmatic example we can think of
the homogeneous $d-$dimensional Poisson point process. In this case 
${\cal P}_N (\{ \bx_i \})=V^{-N}$ simply and independently of the value of $n_0$.
Given a function $X(\{\bx_i\})$ of the $N$ particle positions in thevolume $V$
its average, conditional to the value of $N$, can be written as
\[
\langle X\rangle_N \equiv \int_V \left[\prod_{i=1}^N d^d x_i\right]{\cal P}_N (\{ \bx_i \}) 
X(\{ \bx_i \})\,,
\] 
where the position of each particle is integrated in the volume $V$.
In order to evaluate the {\it unconditional} average of the property
$X$, for which all possible outcomes of the value $N$ are considered,
one would need the probability $q_N$ of having $N$ particles in the
volume $V$, which permits to write:
\be
\langle X\rangle=\sum_{N=0}^{\infty}q_N\langle X\rangle_N\,,
\label{GC-av}
\ee
in a strict analogy with the grand canonical ensemble average in
equilibrium statistical mechanics.  However, since we are restricting
the discussion to large scale uniform particle distributions, for
which $\delta N(R)/\left<N(R)\right>$ vanishes for asymptotically
large $R$, we expect that the larger the volume $V$ the narrower will
be the peak around $N=\left<N(V)\right>= n_0 V$ in which the measure
$q_N$ will be concentrated (for simplicity we have indicated with $V$
both the region and its size). Asymptotically we expect that only the 
term of index $N_0V$ will contribute to the sum in Eq.~(\ref{GC-av}),
i.e., for sufficiently large $V$ we can write:
\[\langle X\rangle\simeq \langle X\rangle_{N_0V}\,.\]
In other words we can consider that for sufficiently large $V$ the
conditional PDF ${\cal P}_{n_0V} (\{ \bx_i \})$ characterizes
completely the statistical properties of the particle distribution in
the finite volume $V$ and use this to evaluate in the following
subsection the statistical properties of the total force. 
This is exactly what has been done, for instance, by
Chandrasekhar in \cite{chandra43} to calculate the total gravitational
force PDF in the Poissonian case.


In Appendix \ref{appendix-SPP} we recall some of the basic
definitions and properties of the statistical characterizations of
uniform SPP. We will use below notably two essential properties 
of $S(\bk)$,  the  {\it structure factor} (SF),  which follow from 
its definition:

\begin{itemize}
\item
\be
\lim_{k \rightarrow 0} k^d S(\bk) =0 \,,
\label{Sk-uniformity}
\ee
i.e, the SF is an integrable function of $\bk$ at
$k=0$, and 
\item
\be
\lim_{k \rightarrow \infty} S(\bk)= 1\,.
\label{SF-kinfty}
\ee
\end{itemize}

\subsection{General expression for the force PDF}

Let us consider now that the particles in any realization of the SPP
interact through a pair force ${\bf f} (\bx)$, i.e., ${\bf f} ({\bf x})$
is the force exerted by a particle on another one at vectorial 
separation $\bx$. Further we will assume that the pair force is 
\begin{itemize} 
\item {\it central}, i.e., 
\be
{\bf f} (\bx)=\hat{\bx}f(x)\,,
\label{f-isotropy}
\ee 
where $\hat{\bx}=\bx/x$, and
\item
{\it bounded}, i.e., there
exists $f_0 < \infty $ such that $|{\bf f} (\bx)|=f(x) \le f_0$ for
all $\bx$. 
\end{itemize} 

These assumptions simplify our calculations considerably, but 
do not limit our aim which is to establish the relation solely between 
the statistical properties of the force field and {\it the behavior of the 
pair interaction at  large distances}. Note that the second
assumption means that, in cases such as the gravitational or 
the Coulomb interaction, the divergence at zero separation is 
assumed appropriately regularized. We will briefly describe
in our conclusions below how our results could be generalized
to include such singularities.

Let us assume for the moment that the system volume $V$ is {\it
 finite}.  As shown above, if $V$ is sufficiently large, one can
consider that the number of particles in this volume is
deterministically $N_0V$.  We will deal with the important problem
of the infinite volume limit defined by $N,V\to \infty$ with $N/V\to
n_0>0$ in the next subsection, by studying directly the limit $V\to
\infty$ with fixed $N_0V$.  The total force field ${\bf F} ({\bf
  x})$ at a point ${\bf x}$, i.e., the force on a test particle placed
at a point $\bx$, may thus be written \be \bF (\bx)=\sum_{i=1}^N {\bf
  f} (\bx-\bx_i)= \sum_{i=1}^N \frac{\bx-\bx_i}{|\bx-\bx_i|}f
(|\bx-\bx_i|)\,.
\label{force-finite-sum}
\ee 
The force field ${\bf F} ({\bf x})$ may be considered
as a stochastic variable with respect to the SPP.  Choosing
arbitrarily the origin as the point where the total force is
evaluated, the PDF of this force is formally defined by\footnote{We
  consider here the {\it unconditional} force PDF, i.e., the force is
  that at an arbitrary spatial point, rather than that on a point
  occupied by a particle which belongs to the particle
  distribution. It is the latter case, of the {\it conditional} force
  PDF, which is often considered in calculations of this kind (see
  e.g. \cite{wesenberg+molmer_PDFdipoles_2004,
    andrea_1dforcesPDF_2005, PDF_SL_2006}).  The distinction is not
  important here as the constraints we derive, which depend on the
  {\it large scale} correlation properties of the particle
  distribution, would be expected to be the same in both cases.}
\[
P_N (\bF)= \int_V \left[\prod_{i=1}^{N}d^d x_i\right] {\cal P}_N(\{\bx_i\}) 
\delta \left[\bF+\sum_i {\bf f} (\bx_i) \right]\,,
\]
where we have used, as assumed, that 
${\bf f} (-\bx_i)=-{\bf f} (\bx_i)$.
Using the identity 
\be
\delta(\by)= \frac{1}{(2\pi)^d} \int d^d q\, e^{i \bq \cdot \by}
\ee
this can be rewritten as
\[
P_N (\bF)= \frac{1}{(2\pi)^d} \int d^d q\, e^{i \bq \cdot \bF}
 \int_V  \left[\prod_{i=1}^{N}d^dx_i\,e^{i \bq \cdot {\bf f} (\bx_i)}\right]\, 
{\cal P}_N (\{\bx_i\}) \,.
\]
The integral over the spatial coordinates in the above equation defines
the {\em characteristic function} of the total field $\bF$
\be 
\tilde P_N(\bq)=\int_V \left[\prod_{i=1}^{N}d^dx_i\,e^{i \bq \cdot {\bf f}
    (\bx_i)}\right]\, {\cal P}_N (\{\bx_i\})\,,
\label{PNqx}
\ee
so that
\[P_N (\bF)= \frac{1}{(2\pi)^d} \int d^d q\, e^{i \bq \cdot \bF}
\tilde P_N(\bq)\,.\]
The integral over spatial configurations in Eq.~(\ref{PNqx})
can be conveniently rewritten as an integral over the
possible values of the pair forces due to each of the $i=1,...,N$
particles: 
\be
\tilde P_N (\bq) \equiv \int \left[\prod_{i=1}^{N}d^d f_i\, 
e^{i \bq \cdot {\bf f}_i} \right]\, {\cal Q}_N
(\{ {\bf f}_i\})\,,
\label{PNqf}
\ee
where 
\be
{\cal Q}_N(\{ {\bf f}_i\})=\int_V \left[\prod_{i=1}^{N}d^d x_i\right] 
{\cal P}_N(\{\bx_i\})\prod_{i=1}^{N}\delta[{\bf f}_i-{\bf f}(\bx_i)] 
\label{P-f-P-x}
\ee 
is the joint PDF for the pair forces ${\bf f}_i$.  Note that,
since $\bF$ is the sum of the variables $\{{\bf f}_i\}$ its
characteristic function $\tilde P_N (\bq)$ can be 
given as 
\be
\tilde P_N (\bq) = \tilde{\cal Q}_N(\{ {\bf q}_i = \bq \})
\label{PN-Qq}
\ee
where $\tilde{\cal Q}_N(\{ {\bf q}_i\})$ is the $Nd-$dimensional FT
of the joint pair forces PDF ${\cal Q}_N(\{ {\bf f}_i\})$, i.e., 
\be
\tilde {\cal Q}_N(\{ {\bf q}_i\})=\int  \left[\prod_{i=1}^{N}d^d f_i 
e^{i \bq_i \cdot {\bf f}_i}\right]\, 
 {\cal Q}_N(\{ {\bf f}_i\}) \,.
\label{Qqi-defn}
\ee

\subsection{Analyticity properties of the force PDF}
\label{Analyticity properties of the force PDF}

From the fact that the pair force is {\it bounded} it follows
that ${\cal Q}_N(\{ {\bf f}_i\})$ has a compact
support, and, since it is absolutely integrable (by definition), 
FT theory (see e.g. \cite{kolmogorov+fomin}) implies that its 
characteristic function 
$\tilde {\cal Q}_N(\{ {\bf q}_i\})$ is an analytic function of the 
variables $\{{\bf q}_i\}$. Consequently  $\tilde P_N (\bq)$
is an analytic function of $\bq$.  Again from FT theory
one has therefore that $P_N (\bF)$ is a rapidly  
decreasing function  of $\bF$: 
\[\lim_{F\to\infty}F^\alpha P_N (\bF)=0\,,\;\;\forall\alpha>0.\] 
Thus  $P_N (\bF)$ is a well-defined function of which  
all moments finite, i.e.,  $0  <  \left<|\bF|^n\right><+\infty$ for any $n\ge 0$.  

{\it Let us now consider what happens when we take the limit}
$V\to\infty$ with $N_0V$. On one hand the joint PDF
${\cal Q}_N(\{ {\bf f}_i\})$ remains non-negative and absolutely
integrable at all increasing $V$. 
On the other hand the support of this function remains
compact with a diameter unaffected by the values of $V$, but
fixed only by $f_0$. Therefore we expect that the FT theorem 
keeps its validity also in the infinite system limit resulting 
in an analytical
\[\tilde P(\bq)
\equiv\lim_{V\to\infty\atop N/V_0}\tilde P_N (\bq)\,.\]
Therefore we will have that
\[P(\bF)\equiv\lim_{V\to\infty\atop N_0V} P_N (\bF)\] 
satisfies
\[\lim_{F\to\infty}F^\alpha P(\bF)=0\,,\;\;\forall\alpha>0.\] 
There are then only two possibilities for the behavior of
$\tilde P_N(\bq)$ in the infinite system limit:

\begin{enumerate}
\item It converges to an absolutely integrable
function which is {\em not identically zero} everywhere, giving  
a $P( \bF )$ which is normalizable and non-negative on its 
support.  Further all the integer moments of $|\bF|$ are 
positive and  finite. 
  
\item  It converges to zero everywhere, giving 
$P(\bF) \equiv 0$. More specifically $P_N (\bF)$ with $N_0V$ converges
point-wise to the null function: it becomes broader and broader 
with increasing $N$ (and $V$), but with an  amplitude which
decreases correspondingly and eventually goes to zero 
in the limit.
  
\end{enumerate}

This latter case is analogous to the case of the sum of identically 
distributed uncorrelated random variables: if this sum is not normalized
with the appropriate power of the number $N$ of such variables, the PDF
 of the sum vanishes point-wise in a similar way in the limit $N\to\infty$.

In summary it follows from these considerations of the analyticity
properties of $\tilde P_N(\bq)$ at increasing $V$ that the case of a
well defined, but fat tailed $P(\bF)$, can be excluded: in
the infinite system limit the force PDF, if defined, is expected to be
a normalizable and rapidly decreasing function.

\section{ Large distance behavior of pair interactions and the force PDF}
\label{PDF-F-mean-variance}

In this section we use the result derived in the previous section
to infer the main result of this paper:  the relation between the 
large scale behavior of the pair interaction and the force PDF
in the infinite system limit.  

We thus consider, as above, a central and bounded pair force
such that
\be
\label{force-gen}
f(x)\simeq \frac{g}{x^{\gamma+1}}\;\;\mbox{for }x\to\infty\,, 
\ee
or, equivalently, a pair interaction corresponding to a two-body 
potential $V(x)\simeq g/(\gamma x^\gamma)$ 
at large $x$ for $\gamma\neq 0$ (and from 
$V(x)\simeq -g\ln x$ for $\gamma=0$).  Since the
pair force is bounded, we have $\gamma > -1$.

Given the final result derived in the previous section,  it follows
that, to determine whether the force PDF exists, it is sufficient
to analyze a single {\it even} moment  of this  PDF:  because
the PDF, when it exists, is rapidly decreasing, any such moment
is necessarily finite and non-zero in this case, and diverges
instead when the PDF does not exist. We choose to analyze 
the behavior of the second moment,  $\left<F^2\right>$, which
is equal to the variance of the PDF since the first moment 
$\left<\bF\right>$ is zero (see below).
We choose this moment
because, as we will now see,  it can be expressed solely 
in terms of the FT of ${\bf f}(\bx)$ and of the SF of the 
microscopic density of the particle distribution.  
From these expressions we can then infer easily
our result.

\subsection{Variance of the force in infinite system limit}

The formal expression of the total
force acting on a test particle (i.e. the force field) at $\bx$ in the
infinite system limit may be written  
\be \bF (\bx)= \int d^d x^\prime
\frac{\bx-\bx^\prime}{|\bx-\bx^\prime|}f  (|\bx-\bx^\prime|)
n(\bx^\prime) 
\label{force-full}
\ee
where the integral is over the infinite space and $n(\bx)$,
given in Eq.~(\ref{density}),  is the density
field in a realization of the general class of {\it uniform} SPP we have 
discussed with positive mean density $n_0$.
 
It is simple to show, using  Eq.~(\ref{force-full}) and 
the definition of the SF given above in Eq.~(\ref{Sk-definition}),
that formally
\be
\langle \bF ^2 \rangle= 
\frac{1}{(2\pi)^d}\int  d^d k |\tilde{{\bf f}} (\bk)|^2 S(k)
\label{force-variance}
\ee 
where $\tilde{{\bf f}} (\bk)$ is the ($d$-dimensional) FT 
of ${\bf {\hat x}} f(x)$. It is straightforward to 
show that $\tilde{{\bf f}} (\bk)= {\bf {\hat k}} \tilde{f} (k)$,
where the explicit expression for $\tilde{f} (k)$ is given 
in the appendix\footnote{Note that only in $d=1$ does 
$\tilde{f}(k)$ coincide with
the direct FT of $f(x)$.}. We can thus write 
\bea
\label{force-variance-isotropic}
\langle \bF ^2 \rangle&=& 
\frac{1}{(2\pi)^d}\int d^d k |\tilde{f}(k)|^2 S(k)\\
&=&\frac{1}{2^{d-1}\pi^{d/2}
\Gamma(d/2)}\int_0^\infty dk\,k^{d-1}|\tilde{f}(k)|^2 S(k) \,,\nonumber
\eea 
where $\Gamma(x)$ is the usual Euler Gamma function.

\subsection{Force PDF for an integrable pair force}
\label{Force PDF for an integrable pair force}

Let us now consider the integrability of the integrand in 
Eq.~(\ref{force-variance-isotropic}).
We start with the case in which $f(x)$ is not only bounded but 
integrable in $\mathbb{R}^d$, i.e., with $\gamma>d-1$.
Given these properties, it is straightforward
to verify, using  the conditions ~(\ref{Sk-uniformity}) and 
(\ref{SF-kinfty}) on $S(k)$ and standard FT theorems, that
the function  $|\tilde{f} (k)|^2 S(k)$) is also 
integrable in $\mathbb{R}^d$. The variance is therefore
finite, from which it follows that the PDF exists,
and furthermore that all its moments are finite.

\subsection{Force PDF for a non-integrable pair forces}
\label{Force PDF for a non-integrable pair forces}

For a pair force which is absolutely non-integrable, i.e., 
$\gamma < d-1$,  the FT $\tilde{\bfo} (\bk)$ of $\bfo(\bx)$ in 
Eq.~(\ref{force-variance-isotropic}) is defined only
in the sense of distributions, i.e., the integrals
over all space of $f(x)$ must be defined by
a symmetric limiting procedure. Physically
this means that the expression Eq. ~(\ref{force-full})
for the force  on a particle in infinite space 
must be calculated as 
\be \bF (\bx)= 
\lim_{\mu \to 0^+} \lim_{V\to\infty}\int_V  \frac{\bx-\bx^\prime}{|\bx-\bx^\prime|} f  (|\bx-\bx^\prime|)
e^{-\mu |\bx-\bx^\prime|}
n(\bx^\prime) d^d x^\prime \,,
\label{force-full-regularised}
\ee 
where the two limits do not commute.  In other words,
  $\bF (\bx)$ is defined as the zero screening limit of a screened
  version of the simple power law interaction in an infinite
  system. The expression Eq.~(\ref{force-variance-isotropic}) is then
meaningful when $\tilde{f} (k)$ is taken to be defined in the
analogous manner with the two limits $\mu \to 0^+$ of the screening and
  $V\to\infty$ (i.e. with the minimal non-zero mode $k\sim 1/V\to 0^+$) taken
  in the same order as indicated in Eq.~(\ref{force-full-regularised}).

Let us consider then again, for the case $\gamma < d-1$,  
the integrability of the integrand in  Eq.~(\ref{force-variance-isotropic}).
To do so we need to examine in detail the small $k$ behavior of
$\tilde{f} (k)$. It is shown in the appendix that,
as one would expect from a simple dimensional analysis, 
for $f(r \rightarrow \infty) \sim 1/r^{\gamma+1}$ we have 
$f (k \rightarrow 0) \sim k^{-d+\gamma +1}$ in any $d$, for 
the case of a pair force which is not absolutely integrable, and 
bounded, i.e., $-1 < \gamma < d-1$. It follows then from 
Eq.~(\ref{force-variance-isotropic}) that the variance is finite
for a given $\gamma$ only for a sub-class of uniform
point processes, specifically those which satisfy 
\be
\lim_{k \rightarrow 0}  k^{-d+2\gamma+2} S(k)=0\,,
\label{asymptotic-FT-f}
\ee  
i.e., for $S(k\rightarrow 0) \sim k^n$ 
with 
\be
n>d-2\gamma-2=-d+2(d-1-\gamma)\,. 
\label{condition-longrange}
\ee
For uniform point processes violating this condition, 
i.e., with $S(k\rightarrow 0) \sim k^n$  and
$-d<n \leq -d+2(d-\gamma-1)$, the variance diverges.
It follows from the results on the PDF of $\bFo$ presented
in the previous section that the total force itself $\bFo(\bx)$ is 
then badly defined in the infinite system limit.


These results of Sec. \ref{Force PDF for an integrable pair force}
and Sec. \ref{Force PDF for a non-integrable pair forces} combined
are the central ones in this paper, anticipated in the introduction.

{\it Firstly},  when pair forces are absolutely integrable at large 
separations, the total force PDF is well defined in the infinite 
system limit, while for pair forces which are not absolutely
integrable this quantity is ill defined. This has the simple
physical meaning anticipated in the introduction: when 
this PDF is well defined, the force on a typical particle 
takes its dominant contribution from particles in a finite 
region around it;  when instead the PDF is ill defined far-away 
contributions to the total force dominate, diverging with the size 
of the system. Thus absolutely integrable pair forces with 
$\gamma > d-1$ are, in this precise sense,  ``short-range", while 
they are ``long-range" when $\gamma  \leq  d-1$. To avoid confusion
with the usual classification of the range of interactions based on 
the integrability properties of the interaction potential, we will adopt
the nomenclature that interactions in the case $\gamma > d-1$
are {\it dynamically short-range}, while for  $\gamma \leq d-1$ they
are {\it dynamically long-range}.  Thus an interaction with 
$d-1 < \gamma \leq  d$ can be described as {\it thermodynamically
long-range but dynamically short-range}. 

{\it Secondly} the results in Sec. \ref{Force PDF for a non-integrable pair forces}
detail how, for $\gamma \leq d-1$,  the force PDF in the infinite system limit may be
defined provided an additional prescription is given for the calculation of 
the force. In the next section we explain the physical meaning and relevance
of this result. 

\section{Definedness of dynamics in an infinite uniform system}
\label{Definedness of dynamics in the infinite system limit}

The regularization Eq.~(\ref{force-full-regularised}) is simply
the generalization to a generic pair force with $\gamma \leq d-1$
of one which is used for the case of Newtonian gravity, 
often referred to as the ``Jeans swindle'' (see e.g.  \cite{binney}).
It was indeed originally introduced by Jeans \cite{jeans} in
his treatment of self-gravitating matter in an infinite universe.
However,  as explained by Kiessling in \cite{kiessling},  its
denomination as  a  ``swindle'' is very misleading, as it can 
be formulated in a mathematically rigorous and physically
meaningful manner, precisely as in Eq.~(\ref{force-full-regularised}). 

The prescription Eq.~(\ref{force-full-regularised}) simply makes
the force on a particle defined by setting to zero the ill defined 
contribution due to the non-zero mean density:
\be \langle \bF (\bx) \rangle = 
\lim_{\mu \to 0^+} n_0 \int  \frac{\bx-\bx^\prime}{|\bx-\bx^\prime|} f  (|\bx-\bx^\prime|)
e^{-\mu |\bx-\bx^\prime|} d^d x^\prime =0 \,,
\label{force-full-regularised-average}
\ee
The force on a particle can thus be written as 
\be \bF (\bx)  = 
\lim_{\mu \to 0^+}  \int  \frac{\bx-\bx^\prime}{|\bx-\bx^\prime|} f  (|\bx-\bx^\prime|)
e^{-\mu |\bx-\bx^\prime|}
\delta n(\bx^\prime) d^d x^\prime \,,
\label{force-full-regularised-fluctuation}
\ee
where $\delta n(\bx^\prime)= n(\bx^\prime) -n_0$ is the density fluctuation
field.  It is straightforward to show that the derived constraint (\ref{condition-longrange}) 
corresponds  simply to that which can be anticipated  by a naive analysis of the 
convergence of the integral Eq.~(\ref{force-full-regularised-fluctuation}):  treating $\delta n(\bx^\prime)$ 
as a deterministic function (rather than a stochastic field) one can require 
it to decay at large $|\bx^\prime|$ with a sufficiently large exponent in order to give integrability; taking
the FT to infer the behavior of  $|\tilde{\delta} n(\bk)|^2$ one obtains the condition
(\ref{condition-longrange}). 

The relevance of the results we have derived for the force PDF in the 
infinite system limit using this regularization arises thus, as it does 
in the case of Newtonian  gravity, when one addresses the following 
question:  is it possible to define consistently {\it dynamics} under 
a given pair interaction in an infinite system which is uniform at large scales? 
As we now discuss, generalizing considerations given by
two of us in  \cite{1dPoisson_2010} for the specific case of
gravity in $d=1$,  the answer to this question
is in fact phrased in terms of the definedness of the PDF of
{\it force differences} rather than that of forces. This leads 
then to our second classification of pair interactions.

\subsection{Evolution of fluctuations and definedness of PDF}
\label{Evolution of fluctuations and definedness of PDF}

Let us consider first an infinite particle distribution which 
is such that the total force PDF is defined {\it at some given time}, i.e.,  
for $\gamma > \ d-1$ we may consider any uniform SSP, while
for  $\gamma < d-1$ we may consider (employing the regularization
discussed) only the class of SSP with fluctuations at large scales obeying 
the condition (\ref{condition-longrange}) {\it at this time}.
The forces on particles at this initial time are then well defined.  This
will only remain true, however, after a finite time interval, if the evolved
distribution continues to obey the same condition (\ref{condition-longrange}). 
Let us determine when this is the case or not.

In order to do so, it suffices to consider the evolution of the 
density fluctuations, 
and specifically of the SF at small $k$, due to the action of this 
force field. Given that we are interested in the long-wavelength modes of the 
density field, we can apply the differential form of the 
continuity equation for the mass (and thus number)
density between an initial time $t=0$ and a time $t=\delta t$:
\be
n(\bx,\delta t) -n(\bx,0) =  {\vec \nabla}[n(\bx,0) \bu (\bx,0)]
\ee
where $\bu (\bx,0)$ is the infinitesimal displacement field.
Subtracting the mean density $n_0$ from both sides, and linearizing
in  $\delta n(\bx,\delta t)=[n(\bx,\delta t)-n_0]$ and $\bu (\bx,0)$, we obtain, on 
taking the FT,
\be
{\tilde \delta n} (\bk,\delta t) ={\tilde \delta n} (\bk,0) 
+ i \, n_0 \, \bk \cdot {\tilde \bu} (\bk,0)\,.
\ee
Taking the square modulus of both sides, in the same approximation 
we get 
\bea
\label{SF-small-k-cnty}
|{\tilde \delta n} (\bk,\delta t)|^2 &-&
|{\tilde \delta n} (\bk,0)|^2 =  \\&&n_0^2 k^2 |\tilde{\bu} (\bk)|^2 
+ 2 \bk n_0 {\rm Im}[ {\tilde \delta n} (\bk,0) \tilde{\bu}^* (\bk, 0)]\,.
\nonumber
\eea
If the displacements are generated solely by the forces acting  
(i.e. assuming velocities are initially zero), we have that
\be 
\bu (\bx, 0) = \frac{1}{2} \bF (\bx, 0) \delta t^2 
\label{displacement}
\ee
and thus, that $|\tilde{\bu} (\bk)|^2 \propto |\bF (\bk)|^2$.
The latter quantity is given, using Eq.~(\ref{force-variance}), by
\be
|\bF (\bk)|^2 = |\tilde{f} (k)|^2 S (k)\,.
\ee
In the analysis in the previous section we used the
result that at small $k$, 
$\tilde{f} (k) \sim k^{-d+\gamma +1}$. Thus 
$|\tilde{\bu} (\bk)|^2 \sim k^{2m+n}$, where 
$m={-d+\gamma +1}$, if $S(k) \sim k^n$. 
It then follows, from
Eq.~(\ref{SF-small-k-cnty}), that the small 
$k$ behavior of the time-evolved SF is given by  
\be
\label{SF-small-k}
S_{\delta t} ( k \to 0) \sim k^n + k^{1+m+n} +k^{2+2m+n}\,.
\ee
It can be inferred that the leading small $k$ behavior of the SF is 
unchanged if and only if $m+1 \geq 0$, i.e., $\gamma \geq d-2$.
Gravity ($\gamma=d-2$) is the marginal case is which
the long wavelength contribution to the SF generated by the evolution has
the same exponent as the initial SF: this
is the well known  phenomenon of {\em linear amplification} of initial
density perturbations (see e.g. \cite{binney, peebles})
which applies\footnote{The result does not 
apply, however, when $n>4$ \cite{peebles};  the reason is 
that fluctuations with $S(k \to 0 ) \sim k^4$  arise generically 
from any rearrangement of matter due to dynamics which conserves 
mass and momentum locally. These effects are neglected
implicitly above when we use the continuum approximation
to the density fluctuation field.}  in infinite self-gravitating 
systems (derived originally by Jeans).

If, on the other hand, $\gamma < d-2$ (i.e. the interaction 
is ``more  long-range'' than gravity in $d$ 
dimensions) the exponent of the small $k$ behavior 
is reduced from $n$ to $n-2(d-2-\gamma)$. Given that our 
result is for an infinitesimal time $\delta t$, this 
indicates in fact a pathological behavior: 
in any finite time interval the exponent $n$ should become,  
apparently,  arbitrarily large and negative, while,
as shown in 
Sect. ~\ref{The force PDF in uniform stochastic point processes},
the constraint $n>-d$ is imposed by the 
assumed large scale uniformity of the SPP.
In other words this result means that, in the infinite system limit, 
when $\gamma < d-2$,  the condition of large scale
uniformity is violated immediately by the dynamical
evolution.  The reason is simply that in this case
the {\it rate of growth of a perturbation at a given scale
increases with the scale}.  Indeed this is the essential 
content of the analysis given just above: 
through the continuity equation, the perturbation to the 
density field is proportional to the {\it gradient of the 
displacement field}, which in turn is simply proportional 
to the {\it gradient of the force}. As we now detail more
explicitly , when $\gamma < d-2$,  this quantity 
diverges with the size of the system.

\subsection{PDF of force differences}
\label{PDF of force differences}

Let us consider now the behavior of the PDF of the difference of the
forces between two spatial points separated by a fixed vector
  distance $\ba$: 
\be 
{\bf {\Delta F}} (\bx; \bx + \ba) \equiv {\bF}
(\bx) -\bF (\bx + \ba)\,. 
\ee 
If this quantity is well defined in the
infinite system limit,  its PDF ${\cal P}({\bf \Delta F}; a)$
will be independent of $\bx$ and will have a parametric dependece
  only on $a=|\ba|$ because of the assumed statistical translational
and rotational invariance of the particle distribution.
 
The analysis of the properties of 
${\cal P}({\bf \Delta F}; a)$ in the infinite volume limit
is formally exactly the same as 
that given above for the total force $\bFo$, with the only replacement
of the pair force in Eq.~(\ref{force-gen}) by the
{\it pair force difference}:
\be
\label{force-gen-diff}
{\bf \Delta f} ( \bx, \bx +\ba ) = {\bf f} ( \bx) - {\bf f} (\bx +\ba )\,, 
\ee
i.e., the difference of the pair forces on two points located at
$\bx$ and $\bx + \ba$ due to a point at the origin. 
Assuming again the possible small scale singularities in this
pair force difference to be suitably regulated, our previous
analysis carries through, the only significant change 
being that, as $x \to \infty$, 
\be
\label{force-gen-diff-infty}
{\bf \Delta f} ( \bx, \bx +\ba ) \sim a {\bf \hat{x}}/x^{\gamma+2}\,. 
\ee
Proceeding in exactly the same manner to analyse 
${\cal P} ( {\bf \Delta F}; a)$, we find that 

\begin{itemize}
\item For $\gamma > d-2$, i.e., if the {\it gradient of the pair force
at fixed $a$ is an absolutely integrable function} of $\bx$ at large 
separations, the PDF ${\cal P} ( {\bf \Delta F}; a)$ is well defined 
in the infinite system limit, and is a rapidly decreasing function 
of its argument for any SPP.  This is true without any regularization.

\item For $\gamma \leq d-2$, on the other hand,  a well defined
PDF may be obtained only by using the regularization
like that introduced above in Eq.~(\ref{force-full-regularised}). 
Therefore the PDF of the force differences then remains well defined, 
i.e., the force difference $\Delta \bFo(\bx;\ba)$ remains finite at all $\bx$, 
only in a sub-class of SPP defined by the constraint
\be
n>d-2\gamma-4=-d+2(d-2-\gamma)\,. 
\label{condition-longrange-mod}
\ee
For the case of gravity $\gamma = d-2$ this coincides  
with the full class of uniform SPP, while for any 
smaller $\gamma$, it restricts to a sub-class 
of the latter.
\end{itemize}

\subsection{Conditions for definedness of dynamics in an infinite system}

Our analysis in Sec.~\ref{Evolution of fluctuations and definedness of
  PDF} of the evolution of density perturbations under the effect of 
  the mutual pair forces gave the sufficient condition $\gamma
\geq d-2$ for the consistency of the dynamics in the infinite system
limit, but with the assumption that the total force PDF was itself
defined. This means that, in the range $d-2 \leq \gamma < d-1$, 
the result derived applies only to the sub-class of infinite uniform
particle distributions in which the large scale fluctuations obey
the condition (\ref{condition-longrange}).  It is straightforward to 
verify, however, that the analysis and conclusions of  
Sec.~\ref{Evolution of fluctuations and definedness of PDF}  can be 
generalized  to cover all uniform SPP for $\gamma \geq d-2$.
In line with the discussion given above, the analysis
requires in fact only assumptions about the behavior of the gradient 
of the forces, rather the forces themselves.  More specifically, the 
only equation which explicitly contains
the force, Eq.~(\ref{displacement}), is a purely formal step which
can be modified to include the possibility that the force diverges
with system size. Indeed if the force --- at a given point ---
includes such a divergence it is sufficient that this divergence
cancels out when we calculate the difference between this force and
that at a neighboring point.  Physically this means simply that, as
discussed above, when we consider the relative motions of particles,
it is sufficient to consider relative forces. Further, as we are
considering the limit of an infinite system in which there is no
preferred point (i.e. statistical homogeneity holds), only relative
motions of points has physical significance, and therefore 
only the spatial variation of the forces can have physical 
meaning. These latter statements can be viewed as a kind 
of corollary to Mach's principle: if the mass distribution of the 
universe is, as it is in the case we consider,  such that there is no 
preferred point in space (and, specifically,  no center of mass) 
inertial frames which give absolute meaning to forces 
(rather than tidal forces) cannot be defined.

In summary our conclusion is that the necessary and
sufficient condition for  dynamics to be defined in the infinite system limit ---
in analogy to how it is defined for Newtonian 
self-gravitating particles in a infinite universe of
constant density --- is that the gradient of the pair
force be absolutely integrable at large separations.
Gravity is the marginal (logarithmically divergent) case 
in which such a dynamics can be defined, but
only by using a prescription such as Eq.~(\ref{force-full-regularised}).
Further these conditions 
on the range of pair forces  can be expressed simply as 
one on the existence of the PDF of force differences
of points as finite separations in the infinite system
limit.

\section{Discussion and conclusions}

In conclusion we make some brief remarks on how the
results derived here relate to previous work in the literature
on force PDFs. In this context we also discuss the important 
assumption we made throughout the article, that the 
pair force considered was {\it bounded}. Finally we return 
briefly to the question of the relevance of the classification
dividing interactions according to the integrability properties
of the pair force, concerning which we have reported
initial results elsewhere \cite{range_PRL2010}.

The first and most known calculation of the force PDF is that of 
Chandrasekhar \cite{chandra43}, who evaluated it for
the gravitational pair interaction  in an infinite homogeneous 
Poisson particle distribution (in $d=3$). This results in
the so-called {\em Holtzmark distribution}, a probability 
distribution belonging to the Levy class (i.e. power law tailed with a
diverging second moment) with $P(\bF)\sim F^{-9/2}$ at large $F$. 
According to our results here, a well defined PDF may be obtained 
for such a force law, which is {\it not} absolutely integrable at large 
separations, only by using a prescription for the calculation of 
the  force in the infinite system limit. In his calculation 
Chandrasekhar indeed obtains the force on a point by 
summing the contributions from mass in {\it spheres} of radius $R$
centered on the point considered, and then taking 
$R \rightarrow \infty$ (with $n_0$ fixed). This prescription
is a slight variant of the one we have employed (following 
Kiessling \cite{kiessling}): instead of the smooth exponential
screening of the interaction, it uses a ``spherical top-hat" 
screening so that the force may be written formally as 
in Eq.~(\ref{force-full-regularised}) with the replacement
of  $e^{-\mu |\bx-\bx^\prime|}$ 
by a Heaviside function $\Theta ( \mu^{-1} -  |\bx-\bx^\prime|)$.
 It is straightforward to verify that the result of Chandrasekhar 
is unchanged if the smooth prescription  Eq.~(\ref{force-full-regularised})
is used instead.
As the Poisson distribution corresponds to an SF 
$S(k\rightarrow 0) \sim k^n$ with $n=0$, the general 
condition (\ref{condition-longrange}) for the existence of the 
PDF we have derived, which gives $n > -1$ for 
gravity in $d=3$,  is indeed satisfied. 
The fact that the
PDF is power-law tailed (and thus {\it not} rapidly decreasing)
arises from the fact that the calculation of Chandrasekhar
does not, as done here, assume that the singularity in
the gravitational interaction is regularized. Indeed 
it is simple to show explicitly \cite{book} that this power law tail 
arises from the divergence in the pair force at zero separation.
This can be done by considering the contribution to the total force 
on a system particle due to its nearest neighbor particle, which turns 
out to have a power law tail identical, both in  exponent and 
amplitude, to that of the full $P(\bF)$.  

Our analysis shows that it is true in general that well defined,
but power-law tailed
force PDFs, can arise only when there are singularities in the 
pair force:
for a bounded force we have seen that the PDF is necessarily
rapidly decreasing when it exists. More specifically, returning
to the analysis of Sec.~\ref{Analyticity properties of the force PDF}, 
it is straightforward to see that the crucial property we used 
of ${\cal Q}_N(\{ {\bf f}_i\})$, that it have {\it compact support},
is no longer valid when the pair force has singularities.
The analyticity properties which lead to a rapidly decreasing
PDF may then not be inferred.  We note that this is true at 
finite $N$, and has nothing to do with the infinite volume limit,
i.e., the appearance of the associated power-law tail arises
from the possibility of having a single particle which give an
unbounded contribution rather than from the combination of
the contribution of many particles which then diverges in
the infinite system limit. The exponent in such a power-law 
tail will depend on the nature of the divergence  at small 
separation.  More specifically, for a central pair force as
considered above and now with a singularity $f(x \to 0) \sim 1/x^{a}$, a 
simple generalization of the analysis for the case of gravity (see \cite{book}) 
of the leading contribution to the total force coming from 
the nearest neighbor particle leads to the conclusion that 
$P (F \to \infty) \sim F^{-d-\frac{d}{a} }$
(where $F = |\bFo|$). This implies that 
the variance diverges (i.e. the PDF becomes fat-tailed)  
for $a > d/2$.  

Force PDFs have been calculated in various other
specific cases. Wesenberg and Molmer \cite{wesenberg+molmer_PDFdipoles_2004}
derived that of forces exerted by randomly distributed dipoles in $d=3$, 
corresponding to a pair force with $\gamma=2$.  According to our
results this is the marginal case in which a summation prescription is
required for the force, and indeed a prescription using spheres, like that used by 
Chandrasekhar for gravity, is employed. We note that \cite{wesenberg+molmer_PDFdipoles_2004} 
focusses on the power-law tails associated with the singularity at zero separation of
the force,  which lead in this case (as can be inferred from the result summarized 
above) to the divergence of the first moment of the force PDF. One of us (AG) has given 
results previously \cite{andrea_1dforcesPDF_2005} for the PDF for 
a generic power-law interaction in $d=1$ for 
$\gamma > -1$ in our notation above. The conditional force PDF is then derived
for the case of an infinite ``shuffled lattice'' of particles, i.e., particles initially
on an infinite lattice and then subjected to {\it uncorrelated} displacements 
of finite variance, and using again, as Chandrasekhar, a ``spherical top-hat" 
prescription for the force summation (for $\gamma \leq 0$, when the 
pair force is not absolutely integrable).  It is simple to show  \cite{book} that 
such a distribution has an SF with $n=2$ at small $k$, and thus the
existence of the force PDF in these cases is again in line with the 
constraint (\ref{condition-longrange}) derived. Power-law tails
are again observed in these cases, and their exponents related
explicitly to the singularity in the assumed power-law force 
at zero separation. 

The calculation of  Chandrasekhar
has been generalized in  \cite{PDF_SL_2006} to the case of 
particles on an infinite shuffled lattice. This leads again, in line with 
condition (\ref{condition-longrange}), to a well defined PDF,
again with or without power-law tails according to whether the 
singularities in the pair force are included or not. Chavanis
\cite{chavanis-forcePDF-poisson_2008} considers, on the 
other hand, the generalization of Chandrasekhar calculation
(for the PDF of gravitational forces in a Poisson distribution)
to $d=2$ and $d=1$. The condition  (\ref{condition-longrange} 
for gravity ($\gamma=d-2$) gives $n > -d+2$, which implies
that the force PDF is not defined in the infinite system limit 
we have considered for $d \leq 2$, and indeed in 
\cite{chavanis-forcePDF-poisson_2008} well defined 
PDFs are obtained in $d=2$ and $d=1$ by using a
different limiting procedure involving in each case
an appropriate rescaling of the coupling with $N$.  
The physical meaning of such a procedure is discussed 
in \cite{1dPoisson_2010},  which considers in detail the 
calculation of the force PDF for gravity in $d=1$ in a 
Poisson distribution (as in \cite{chavanis-forcePDF-poisson_2008}). 
An exact calculation of the force PDF of the {\it screened}
gravitational force in the infinite system limit is given, which 
allows one to see in this case exactly how the general result
given here is verified in this specific case: all moments 
of the PDF diverge simultaneously as the screening length
is taken to infinity, giving a PDF which converges 
point-wise to zero.  The force PDF for gravity in $d=1$ for a class of infinite
particle distributions generated by perturbing a lattice 
has been derived recently by three of us in  \cite{1dgrav-sl}. 
It is straightforward to show that one of the conditions imposed
on the perturbations to obtain the PDF, that the variance of the perturbations
be finite, corresponds in fact to the condition $n>1$ which coincides 
precisely with the more general condition (\ref{condition-longrange}) derived 
here. Unlike in the other specific cases just discussed, it turns
out that in this case (gravity in $d=1$) it is in fact necessary to
use the smooth prescription  Eq.~(\ref{force-full-regularised}). 
As explained in detail  in \cite{1dgrav-sl}, the top-hat prescription
does not give a well defined result in this case, because 
surface contributions to the force which do not decay with
distance in this case are not regulated by it. We underline
that the general result given in the present article are for this specific 
prescription Eq.~(\ref{force-full-regularised}). Further
analysis would be required to derive the general conditions
in which a top-hat prescription also gives the same 
(and well-defined) PDF. 
 
Finally let us comment on why we anticipate
the classification of pair interactions according to their ``dynamical range'', 
formalized here using the force PDF, should be a useful and relevant one 
physically in the study of systems with long-range interactions. The reason is 
that this classification reflects, as we have explained, the relative importance
of the mean field contribution to the force on a particle, due to the 
bulk, compared with that due to nearby particles. Now it is precisely 
the domination by the former which is understood to give the 
regime of {\it collisionless} dynamics which is expected to lead to the formation 
of QSS states, which are usually interpreted to be stationary states of the Vlasov 
equations describing such a regime of the dynamics (see e.g. \cite{balescu}). 
In a recent article \cite{range_PRL2010} by three of us, we have reported a numerical
and analytical  study which provides strong evidence for the
following result, very much in line with this naive expectation:  systems of particles 
interacting by attractive power law pair interactions like those considered 
here can always give rise to QSS; however when the pair force is {\it dynamically
short-range} their existence requires the presence of a sufficiently large soft 
core, while in the {\it dynamically long-range} case QSS can occur
independently of the core, whether hard or soft, provided it is sufficiently
small. In other words only in the case of a pair force which 
is ``dynamically long-range" can the occurrence of QSS be considered 
to be the result only of the long distance behavior of the interaction alone.
This finding is very consistent with what could be anticipated 
from the preceding (naive) argument: the effect of a ``soft core''
is precisely to reduce the contribution to the force due 
to nearby particles, which would otherwise dominate over
the mean field force in the case of a pair force which is
absolutely integrable at large distances. Indeed the
meaning of  ``sufficiently large'' specified in \cite{range_PRL2010} 
is that the size of the soft core must increase in an appropriate 
manner with the size of the system as the limit $N \rightarrow \infty$ 
is taken, while we have always implicitly assumed it to be fixed
in units of the interparticle distance here. Further work on
these issues will be reported elsewhere.

\vskip 0.5 cm

We thank M. Kiessling and T. Worrakitpoonpon for useful 
conversations, and P. Viot for useful comments on the manuscript.  

\appendix

\section{One and two point properties of uniform SPP}
\label{appendix-SPP}

In this appendix we give the general one and two-point statistical 
characterization of a SPP which is uniform on large scales.

The description of the correlation properties of a generic
uniform SPP is given by the $n$-point correlation
functions of the density field. For our considerations
it will turn out to be sufficient to consider only
the two-point properties, and more specifically
it will be most convenient to characterize them
in reciprocal space through the {\it structure factor}
(SF) (or power spectrum). This is defined by 
\begin{equation}
\label{Sk-definition}
S({\bf k}) = \lim_{V \rightarrow \infty}
\frac{\left<|\tilde{\delta} n ({\bf k}; V)|^2\right>}{n_0 V}
\end{equation}
where
\begin{equation}
\tilde{\delta} n ({\bf k}; V) = \int_V d^d x\, e^{-i {\bf k}\cdot{\bf x}}
[n ({\bf x}) - n_0]\,.
\end{equation}
With these normalisations the SF of an uncorrelated
Poisson process is $S(\bk)=1$. For a statistically
isotropic point process $S(\bk)\equiv S(k)$, where
$k=|\bk|$. We recall here that $S(\bk)$ is the Fourier transform (FT)
of the connected two point density correlation function:
\[S(\bk)=\int d^d x\, e^{-i {\bf k}\cdot{\bf x}}C(\bx)\]
where
\[C(\bx)=\frac{\left<n(\bx_0+\bx)n(\bx_0)\right>-n_0^2}{n_0}=
\delta(\bx)+n_0h(\bx)\,.\] In the last expression we have explicitly
separated in the correlation function $C(\bx)$ the shot noise term
$\delta(\bx)$, present in all SPP and due to the ``granularity'' of
the particle distribution, from the ``off-diagonal'' term $n_0h(\bx)$
which gives the actual spatial correlations between different
particles.

In the paper we study the convergence properties of forces at large
distances and are thus mainly interested in the properties of the SF at
small $k$. In this respect we will use the following limit on the SF
which follows from the assumed uniformity of the SPP: 
\be
\nonumber
\lim_{k \rightarrow 0} k^d S(\bk) =0 \,,
\ee
i.e, the SF is an integrable function of $\bk$ at
$k=0$. This constraint simply translates in
reciprocal space the requirement from uniformity
on the decay of relative fluctuations of the number 
of particles contained in a volume $V$ about the mean
at large $V$:
\be
\nonumber
\lim_{V \to \infty} 
\frac{\langle N(V)^2 \rangle - \langle N(V) \rangle^2}   
{\langle N(V) \rangle^2}   =0 \,.
\ee
Given that $\langle N(V) \rangle \propto V$,
the root mean square fluctuation of particle number $N$ in a 
volume $V$ must diverge slower than the volume $V$ itself
in order that this condition be fulfilled.
(This is equivalent to saying 
that $C(\bx)$ must vanish at large $x$).

We use likewise in the paper only one constraint on the large 
$k$ behavior of the SF, which is valid for any uniform
SPP (see e.g. \cite{book}) and coincides with the shot noise 
term in the correlation function $C(\bx)$:
\be
\nonumber
\lim_{k \rightarrow \infty} S(\bk)= 1\,.
\ee

\section{Small k behavior of $\tilde{\textbf{f}}(\textbf{k})$}
\label{appendix-kbehavior}

We are interested in the small $k$ behavior of the Fourier transform
 $\tilde{\textbf{f}}(\textbf{k})$ of the pair force in $d$ dimensions in 
the case where the pair force $\textbf{f}(\textbf{x}) 
= \hat{\textbf{x}}f(x)$, where $\hat{\textbf{x}} = \frac{\textbf{x}}{\vert \textbf{x} \vert}$, 
is non-integrable but converges to zero at $x \to \infty$, i.e., 
 $f(r) \sim x^{-(\gamma+1)}$ at large $x$ with $-1 < \gamma \leq d-1$.\\

We first show that for a function $\textbf{f}(\textbf{x}) 
= \hat{\textbf{x}}f(x)$, its Fourier transform, $\tilde{\textbf{f}}(\textbf{k})=\textrm{FT}[\bf{f}(\bf{x})](\textbf{k})$, 
can be written 
$\tilde{\textbf{f}}(\textbf{k}) = \hat{\textbf{k}}~\psi(k)$ where $\psi(k)$ is 
a function depending only on the modulus of $\textbf{k}$ and 
$\hat{\textbf{k}} = \frac{\textbf{k}}{\vert \textbf{k} \vert}$.  In order to obtain
 this result, we start by writing
\begin{equation}
 \nonumber
 \tilde{\textbf{f}} (\textbf{k}) = \int d^dx~\textbf{f} (\textbf{x}) e^{-i\textbf{k.x}} 
= \int d^dx~\hat{\textbf{x}} f(x)e^{-i\textbf{kx}} ~,
\end{equation}
where this integral is defined in the sense of functions or distributions according to 
the integrability of $f(x)$.\\
In the following we denote by $(\bf{\hat{e}_1}, \bf{\hat{e}_2}, \dots, \bf{\hat{e}_n})$ 
the cartesian vector basis in $d$-dimension and we define 
$(r, \theta_1, \theta_2, \dots, \theta_{d-1})$ the hyper-spherical coordinates of \textbf{x}.
 Considering  $\textbf{k} = k~\bf{\hat{e}_1}$ and denoting for simplicity $\theta = \theta_1$, 
we can write
\begin{equation}
\nonumber
 \tilde{\textbf{f}} (\textbf{k})= \int d^dx~\hat{\textbf{x}} f(x) e^{-ikx cos\theta}~,
\end{equation}
where
\be
\nonumber
d^d x = \Big(\prod_{j=0}^{d-1} \sin^j(\theta_{d-j}) d\theta_{d-j}\Big) x^{d-1}dx\,.
\ee
Projecting $\tilde{\textbf{f}} (\textbf{k})$ on the cartesian basis, 
it is easy to see that the only non-vanishing term is 
$\bf{\hat{e}_1}.\tilde{\textbf{f}} (\textbf{k})$ which gives
\begin{eqnarray}
\nonumber
&&\hat{e}_1 .\tilde{f} (\textbf{k}) = 
C_{\theta_{i\neq 1}} \int_{0}^{\infty}dx x^{d-1} \\
&&\times\int_{0}^{\pi} d\theta \sin^{n-2}(\theta) \cos\theta
f(x) e^{-ikx cos\theta}~,\nonumber
\end{eqnarray}
where $C_{\theta_{i\neq 1}}$ is a constant term coming from the integration over 
all the hyper-spherical coordinates $\theta_i$ with $i \neq 1$. We thus can write
 $\tilde{\textbf{f}} (\textbf{k}) = \hat{\textbf{k}}~\psi(k)$ where $\psi(k)$ is 
a function depending only on the modulus of $\textbf{k}$.\\

We now focus our attention on the small $k$ behavior of the term
\begin{equation}
\label{integral}
 \int_{0}^{\infty} dx x^{d-1} f(r) e^{-ikx cos\theta}~,
\end{equation}
where the function $f(x)$ is non-integrable but converges to zero at $x \to \infty$, i.e.,
 $f(x) \sim x^{-(\gamma+1)}$ at large $x$ with $-1 < \gamma \leq d-1$, 
and thus can be written $f(x) = x^{-(\gamma+1)} + h(x)$ 
with $h(x)$ a smooth function, integrable at $x=0$ and such that 
$x^{\gamma +1} h(x) \to 0$ for $x \to \infty$.\\
Defining explicitly eq.(\ref{integral}) in the sense of distributions, the small $k$ behavior is determined by this leading divergence at $x \to \infty$,
\begin{equation}
\label{screening}
 \lim_{\mu \to 0} \int_{0}^{\infty}dx\, x^{d-1} \frac{e^{-\mu x}}{x^{\gamma+1}}e^{-ikx\cos\theta} ~,
\end{equation}
where the parameter $\mu >0$. We define $\alpha = d-\gamma-2$ 
which satisfies $-1 \leq \alpha < d-1$ and rewrite eq. (\ref{screening})
\begin{equation}
 \nonumber
 \lim_{\mu \to 0} \int_{0}^{\infty}dx\, x^{\alpha} e^{-(ik\cos\theta+\mu) x} ~.
\end{equation}
This can be easily calculated with Laplace's transform and gives
\begin{equation}
 \nonumber
 \int_{0}^{\infty}dx\, x^{\alpha} e^{-(ik\cos\theta+\mu) x} = 
\frac{\Gamma(\alpha+1)}{(\mu + ik\cos\theta)^{\alpha+1}}~.
\end{equation}
We can conclude that 
\begin{eqnarray}
\nonumber
&& \lim_{\mu \to 0} \int_{0}^{\infty}dx x^{d-1} \frac{e^{-\mu x}}{x^{\gamma+1}}e^{-ikx\cos\theta} \\
 &&= i^{-(\alpha+1)} \cos^{-(\alpha+1)}(\theta) \Gamma(\alpha+1) k^{-(\alpha+1)} \sim k^{\gamma-d+1}\,.\nonumber 
\end{eqnarray}
%


\end{document}